\documentclass[thmsa,12pt]{article}
\usepackage{sw20lart}

\input tcilatex
\QQQ{Language}{
American English
}

\begin{document}

\title{PECULIARITIES OF INCLINED FORCE AFFECTION ON HOMOGENEOUS ONE-DIMENSIONAL
ELASTIC LUMPED-PARAMETERS LINE}
\author{S.B.Karavashkin \and Special Laboratory for Fundamental Elaboration SELF,
Head \\
187 apt., 38 bldg., Gagarin Avenue, Kharkov 61140, Ukraine\\
phone 38 (0572) 276624; e-mail sbkarav@altavista.com}
\date{June 1, 2000}
\maketitle

\begin{abstract}
In this paper are analysed exact analytic solutions for one-dimensional
elastic lumped-parameters line under affection of external force inclined to
the axe of line. It is shown that in this case an inclined wave being
drscribed by implicit function propagates along the line. The conclusions
are extended to unforced vibration in the line and to distributed parameters
line vibration. Solution in the form of implicit function is proved as a
generalizing one for the wave equation.
\end{abstract}

\section{Introduction}

In papers [1,2] have been considered longitudinal vibrations in
one-dimensional elastic lumped-parameters line. However in such line
vibrations of more general form are possible, if we attribute the idea of
one-dimensionality only to general shape of line, not to degree of freedom
of vibrating elements of line.

This paper investigates the peculiarities of exact analytic solutions for
this type of problems.

\section{Vibrations in semi-infinite elastic line under inclined external
force affecting on start of line}

Regarding stated above, supposing two degrees of vibration freedom for
elastic line elements, the model can be presented in form shown in fig.1. In
case when amplitude of vibration is small (linear vibration), this model can
be described by two sets of equations - with respect to $x$- and $y$%
-projections of external force affection correspondingly: 
\begin{equation}
\left\{ 
\begin{array}{c}
m\frac{d^2\Delta _1}{dt^2}=F\left( t\right) \cos \alpha +s\left( \Delta
_2-\Delta _1\right) \\ 
m\frac{d^2\Delta _2}{dt^2}=s\left( \Delta _3+\Delta _1-2\Delta _2\right) \\ 
.................................................... \\ 
m\frac{d^2\Delta _n}{dt^2}=s\left( \Delta _{n+1}+\Delta _{n-1}-2\Delta
_n\right) \\ 
...................................................
\end{array}
\right.  \label{1}
\end{equation}
\begin{equation}
\left\{ 
\begin{array}{c}
m\frac{d^2y_1}{dt^2}=F\left( t\right) \sin \alpha +s\left( y_2-y_1\right) \\ 
m\frac{d^2\Delta y}{dt^2}=s\left( y_3+y_1-2y_2\right) \\ 
.................................................... \\ 
m\frac{d^2y_n}{dt^2}=s\left( y_{n+1}+y_{n-1}-2y_n\right) \\ 
...................................................
\end{array}
\right.  \label{2}
\end{equation}
where $\alpha $ is angle of external force inclination to the axe of line.

Each of these sets of equations is similar to ones having been investigated
in [1]. consequently, we can write at once exact analytic solutions for
every of them.

For $x$-component of vibration: periodic regime $\left( \beta <1\right) $%
\begin{equation}
\Delta _n=-j\frac{F_0\cos \alpha }{\omega \sqrt{sm}}\;e^{j\left[ \omega
t-\left( 2n-1\right) \tau \right] }  \label{3}
\end{equation}
aperiodic one $\left( \beta >1\right) $%
\begin{equation}
\Delta _n=\left( -1\right) ^n\frac{F_0\cos \alpha }{\omega \sqrt{sm}}%
\;\gamma ^{2n-1}\;e^{j\omega t}  \label{4}
\end{equation}
critical one $\left( \beta =1\right) $%
\begin{equation}
\Delta _n=\left( -1\right) ^n\frac{F_0\cos \alpha }{2s}\;e^{j\omega t}
\label{5}
\end{equation}
For $y$-component of vibration we obtain correspondingly: periodic regime $%
\left( \beta <1\right) $%
\begin{equation}
y_n=-j\frac{F_0\sin \alpha }{\omega \sqrt{sm}}\;e^{j\left[ \omega t-\left(
2n-1\right) \tau \right] }  \label{6}
\end{equation}
aperiodic one $\left( \beta >1\right) $%
\begin{equation}
y_n=\left( -1\right) ^n\frac{F_0\sin \alpha }{\omega \sqrt{sm}}\;\gamma
^{2n-1}\;e^{j\omega t}  \label{7}
\end{equation}
and critical one $\left( \beta =1\right) $%
\begin{equation}
y_n=\left( -1\right) ^n\frac{F_0\sin \alpha }{2s}\;e^{j\omega t}  \label{8}
\end{equation}

As a result of superposition, there forms an inclined wave propagating in
positive direction of axe $x$; this is confirmed by diagram of vibration
shown in fig.2.

It is characteristic that inclined pattern of vibration remains as under
unforced vibrations in lumped-parameters line as under limiting process to
distributed-parameters line.

Really, basing on results presented in [1], solution, e.g. for unforced
vibration, has the following form: for $x$-component of vibration 
\begin{equation}
\Delta _n=\frac{X_k\cos \left( 2i-1\right) \tau }{\cos \left( 2k-1\right)
\tau }\;e^{j\omega t}  \label{9}
\end{equation}
for $y$-component 
\begin{equation}
y_n=\frac{Y_k\cos \left( 2i-1\right) \tau }{\cos \left( 2k-1\right) \tau }%
\;e^{j\omega t}  \label{10}
\end{equation}
where $X_k$ and $Y_k$ are $x$- and $y$-components of vibration amplitude of $%
k$th element which parameters are specified; $k$ is number of element which
vibration is specified.

In case of limiting process to distributed-parameters line, we can present 
\[
\rho =\frac ma\;;\;\;\;s=\frac Ta\;;\;\;\;n=\frac{x_0}a 
\]
where $\rho $ is density; $T$ is tension in line; $x_0$ is distance from
start of line to the point of rest of investigated element of line; $m$ is
mass of element of line. With it solutions (3)$\div $(8) transform to the
set of equations 
\begin{equation}
\left\{ 
\begin{array}{c}
x=-j\frac{F_0\cos \alpha }{\omega \sqrt{T\rho }}\;e^{j\omega \left( t-x_0%
\sqrt{\rho /T}\right) }+x_0 \\ 
y=-j\frac{F_0\cos \alpha }{\omega \sqrt{T\rho }}\;e^{j\omega \left( t-x_0%
\sqrt{\rho /T}\right) }
\end{array}
\right.  \label{11}
\end{equation}

Obtained set of equations describes paramertically the inclined wave
propagating along the axe $x$, as shown in fig.3.

We can see of carried out investigation that inclined vibrations arise far
from always as a consequence of nonlinear processes in elastic system, as it
was supposed before. Inclined waves can arise quite naturally under
affection of force inclined to the direction of wave process propagation.
And this conclusion can be quite simply extended to a most wide spectrum of
vibration process.

\section{Elements-of-line motion trajectory}

Paying attention to a separate element motion trajectory, we can see easy,
this trajectory has form of ellipse circumscribed around the point of rest
of element. And inclination of wave forms, at the cost of shift phase of
motion along element-to-element elliptic trajectories. Presented structure
of vibration is well-known in physics, particularly in wave processes in
unbounded volumes of liquid. ''In a wave, motion of liquid is
non-stationary. So trajectories of separate particles are far from
coinciding with lines of current in time. They have absolutely other form.
Under small amplitudes they are circumferences in a great approximation. We
find these circular trajectories as on surface as in depth of liquid. Only
in the most upper layers the diameters of circular ways are the most large''
[3, pp.300-301].

Indeed, vibration processes in space have their peculiarities. None the
less, it is characteristic that basic regularities are run down already in
one-dimensional model. It also follows of obtained solutions that
ellipsoidal pattern of vibrations of elements remains as in critical as in
aperiodic regimes. Consequently, in last case in the line forms compound
wave fast-decaying along the line, and this is one more peculiarity that
exact analytic solutions demonstrate.

\section{On new class of functions being the solution of wave equation}

Foregoing generalizations can be extended also to solution of wave equation
in the whole.

It is known that differential equation of hyperbolic type 
\begin{equation}
\frac{\partial ^2y\left( x,t\right) }{\partial x^2}=\frac{k^2}{\omega ^2}\;%
\frac{\partial ^2x\left( x,t\right) }{\partial t^2}=0  \label{12}
\end{equation}
has general solution [5, p.300] 
\begin{equation}
\Phi \left( x,t\right) =\Phi _1\left( kx-\omega t\right) +\Phi _2\left(
kx+\omega t\right)   \label{13}
\end{equation}
where $c=\omega /k$ is velocity of wave propagation, i.e. in the form of two
explicit functions with respect to $\left( x-ct\right) $ and $\left(
x+ct\right) $ correspondingly. Till now this solution was considered the
only and complete, due to theorem of uniqueness of solution of differential
equation. None the less, there exists one more class of functions being the
solution of differential equation (12) but not taken into account by
solution (13). We can present general form of this class of functions in the
form 
\begin{equation}
y\left( x,t\right) =\Phi _1\left( kx-\omega t+\psi _1\left( y\right) \right)
+\Phi _2\left( kx+\omega t+\psi _2\left( y\right) \right)   \label{14}
\end{equation}
where $\psi _1\left( y\right) $and $\psi _2\left( y\right) $are some
twice-differentiable functions. In other words, given solution (14) belongs
to the class of implicit functions whose regularities of behavior and
technique of differentiation and integration essentially differ from such
for explicit functions. Important that, while for explicit functions
definite systematization of differential equations has been created and for
definite class of these equations the regularities and schemes to obtain
solutions have been defined, for implicit functions all these developments
are absent. Naturally, for today correspondence of expression (14) to
differential equation (12) can be checked only by the most simple way - by
straight substitution (14) into (12).

For it, on the grounds of known laws of implicit functions differentiation,
find first and second particular derivatives of expression (14) with respect
to $x$ and $t$. To simplify calculation, consider a half of right part of
expression (14) 
\begin{equation}
y\left( x,t\right) =\Phi _1\left( kx-\omega t+\psi _1\left( y\right) \right)
=\Phi _1\left( A\right)  \label{15}
\end{equation}
where $A=kx-\omega t+\psi _1\left( y\right) $. First derivatives have the
form 
\begin{equation}
\frac{\partial y}{\partial x}=\frac{k\frac{d\Phi _1}{dA}}{1-\frac{d\psi _1}{%
dy}\;\frac{d\Phi _1}{dA}}\;;\;\;\;\frac{\partial y}{\partial t}=\frac{\omega 
\frac{d\Phi _1}{dA}}{1-\frac{d\psi _1}{dy}\;\frac{d\Phi _1}{dA}}  \label{16}
\end{equation}
Second derivatives after transformation and substitution of expressions (16)
take form 
\begin{eqnarray}
\frac{\partial ^2y}{\partial x^2} &=&\frac{k^2\left[ \frac{d^2\Phi _1}{dA^2}+%
\frac{d^2\psi _1}{dy^2}\left( \frac{d\Phi _1}{dA}\right) ^3\right] }{\left(
1-\frac{d^2\psi _1}{dy^2}\;\frac{d\Phi _1}{dA}\right) ^3}  \label{17} \\
\frac{\partial ^2y}{\partial t^2} &=&\frac{\omega ^2\left[ \frac{d^2\Phi _1}{%
dA^2}+\frac{d^2\psi _1}{dy^2}\left( \frac{d\Phi _1}{dA}\right) ^3\right] }{%
\left( 1-\frac{d^2\psi _1}{dy^2}\;\frac{d\Phi _1}{dA}\right) ^3}  \nonumber
\end{eqnarray}
Substituting (17) into (12), we obtain required. Similarly we can prove the
correspondence of second part of expression (14) to equation (12).

Thus, solution (17) defines a whole class of implicit functions satisfying
the linear wave equation. And presence of new class of functions being the
solution of equation (12) does not violate a least the theorem of uniqueness
of solution of differential equation, because under definite conditions 
\[
\psi _1\left( y\right) \equiv 0\;;\;\;\;\psi _2\left( y\right) \equiv 0 
\]
expression (14) degenerates into (13). hereby it is proved that solution
known before is a particular case of more general class of functions.

The found class of implicit functions defines nonlinear wave; its degree of
deformation depends on form of functions $\psi _1\left( y\right) $ and $\psi
_2\left( y\right) $. For example, in particular case of expression (14) (see
fig.4) 
\begin{equation}
y=c\sin \left( kx-\omega t+y\cot \alpha \right)  \label{18}
\end{equation}
solution of wave equation (12) describes progressive wave propagating along
axe $x$ and inclined by angle $\alpha $, what completely corresponds with
inclined vibration in one-dimensional line investigated above.

\section{Conclusions}

1. Analyzing solutions for semi-infinite model on free end of which affects
a force inclined to the axe, we have found that as a result of this
affection, in the line propagate inclined waves described by implicit
function.

2. Solutions of wave equation in the form of implicit functions are
generalizing for known solution being superposition of running waves.

3. We have ascertained that under affection of inclined force the elements
of line follow elliptic trajectories.

\section{Symbols}

$F\left( t\right) $ is external force affecting on the line; $F_0$ is
amplitude of external force; $T$ is tension in the line; $X_i\,$, $Y_i$ are $%
x$- and $y$-components of vibration amplitude of $k$th element whose
parameters of vibration are given; $a$ is distance between the elements of
line; $f$ is frequency of vibration in the line; $i$, $k$, $n$ are indexes; $%
m$ is mass of element of line; $k$ is wave number; $s$ is stiffness
coefficient of line; $t$ is time parameter; $x_0$ is distance from start of
line to the point of rest of last element of line; $y_i$ is displacement of $%
i$th element of elastic line in vertical plane; $\Delta _i$ is instantaneous
longitudinal displacement of $k$th element of line; $\alpha $ is angle of
affecting force inclination to the axe of elastic line; $\beta $, $\gamma
_{+}\,$, $\gamma _{-}\,$, $\tau $ are parameters of line; $\omega $ is
circular frequency of affecting force.

\end{document}